\documentclass[apl,twocolumn,showpacs,amsmath,letter]{revtex4}
\usepackage{graphicx}
\newcommand{\Journal}[4]{#1 \textbf{#2}, #3 (#4)}

\begin{document}

\title{Current-induced reversal in magnetic nanopillars passivated by silicon}
\author{Sergei Urazhdin}
\author{Phillip Tabor}
\affiliation{Department of Physics, West Virginia University,
Morgantown, WV 26506}


\begin{abstract}
We demonstrate that magnetic multilayer nanopillars can be efficiently protected from oxidation by coating with silicon. Both the protected and the oxidized nanopillars exhibit an increase of reversal current at cryogenic temperatures. However the magnetic excitation onset current increases only in the oxidized samples. We show that oxidized nanopillars exhibit anomalous switching statistics at low temperature, providing a simple test for the quality of magnetic nanodevices.
\end{abstract}

\maketitle

Current-induced magnetic switching (CIMS) in magnetic nanopillars caused by spin transfer (ST) from the polarized current $I$ to the magnetic moments is an important mechanism for the operation of nanomagnetic devices.~\cite{slonczewski96} The ability to fabricate nanopillars with reproducible, well-controlled geometry and magnetic properties is critical for the understanding of the current-induced behaviors, and for efficient device applications. 

\begin{figure}
\includegraphics[width=3.2in]{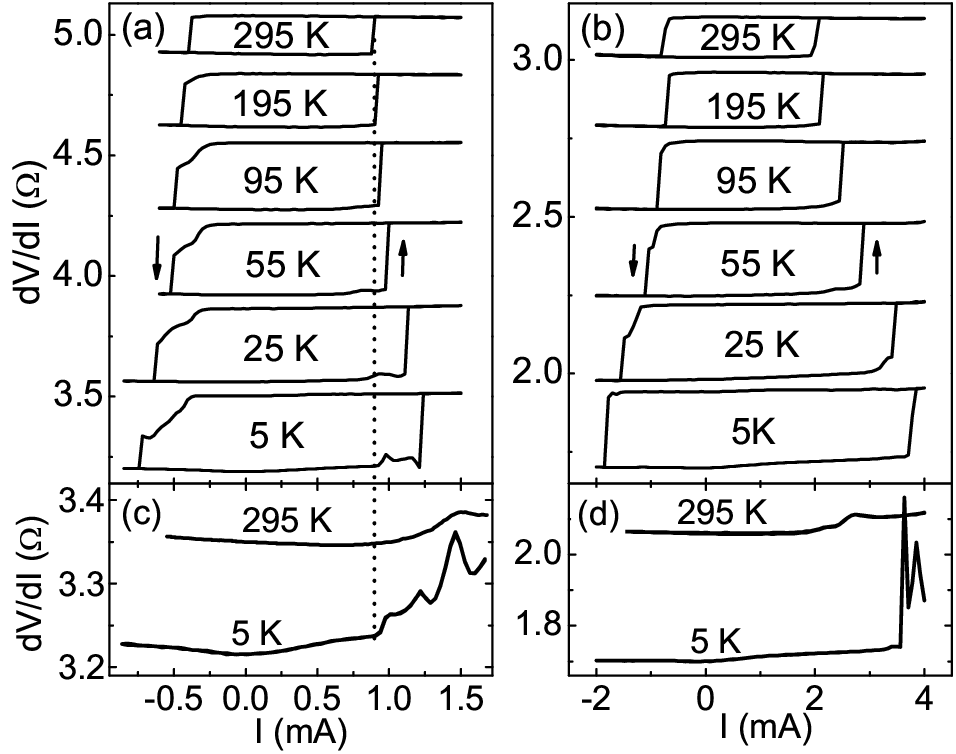}
\caption{\label{fig1} (a,b) $dV/dI$ for samples $A$, $B$ at $H=20$~Oe and the labeled values of $T$. Curves are offset for clarity. Arrows show scan directions. (c,d) same as (a,b), at $H=500$~Oe.}
\end{figure}

Oxidation of the usual Permalloy=Py=Ni$_{80}$Fe$_{20}$ nanopillars  can result in reduced thermal stability at room temperature 295~K (RT), and increased magnetic damping at low temperatures.~\cite{emley,oxidation} Both effects are undesirable for magnetic memory applications, which require a combination of high stability and small damping to facilitate CIMS. To eliminate oxidation, protective coating of nanopillars with AlO$_x$ was developed.~\cite{oxidation} This procedure requires highly precise oxidation of Al to avoid partial shorting of nanopillars by unoxidized Al, or over-oxidation leading to formation of magnetic oxides. Because of these difficulties, it is desirable to develop alternative protection techniques.

Here, we describe a fabrication procedure involving protection of nanopillars with silicon, also serving as an insulating layer between sample leads. Our procedure is simpler than the one involving AlO$_x$, and completely eliminates both the exposure of the nanopillar to oxygen and the possible shunting through the coating. It is thus suitable not only for metallic nanopillars, but also for higher-resistance tunnel junctions. We present measurements of CIMS demonstrating the effectiveness of our protection procedure, relate our results to the  published measurements, and discuss their implications for the mechanism of CIMS.

Multilayers Py(1)Cu(50)Py(20)Cu(8)Py(5)Au(25), where thicknesses are in nm, were deposited on oxidized silicon at RT by sputtering at base pressure of $5\times 10^{-9}$~Torr, in $5$~mTorr of purified Ar. Lithographically patterned  $100$~nm$\times 50$~nm Al nanopillar served as a mask for the subsequent Ar ion milling, which removed the  multilayer down to the middle of the Cu(8) spacer, leaving the Py(20) polarizer unpatterned with dimensions of several micrometers.  For sample type $A$, a $30$~nm-thick undoped Si layer was then sputtered without breaking the vacuum. For sample $B$, $30$~nm of SiO$_2$ was deposited by reactive Si sputtering in $5$~mTorr of Ar:O$_2$ 80:20 mixture. We will show below that this procedure had an oxidizing effect similar to exposure to air, while avoiding surface contamination and variations of atmospheric conditions. 

Subsequent ion milling at $3^\circ$ with respect to the surface removed the insulating cap from the nanopillar. Finally, a $120$~nm thick top Cu lead was deposited after etching in 1:10:1000 solution of HNO$_3$:HF:H$_2$O to clean the nanopillar surface. This cleaning step was more efficient for samples $B$, resulting in somewhat lower resistance. However, the magnetoresistances (MR) of the two sample types were similar. Test samples with a Si spacer but no pillar yielded resistances exceeding $2$~k$\Omega$, suggesting that $30$~nm thick Si is sufficiently insulating not only for metallic nanopillars, but also for tunnel junctions. In contrast to the more common oxide insulators, such as SiO or SiO$_2$, Si spacers do not form pinholes. Lock-in measurements of differential resistance $dV/dI$ were performed in a pseudo four-probe geometry, by superimposing a $20$~$\mu$A rms ac current on the dc current $I$. Magnetic field $H$ was applied along the nanopillar easy axis. Two samples of each type were tested with similar results.

\begin{figure}
\includegraphics[width=3.2in]{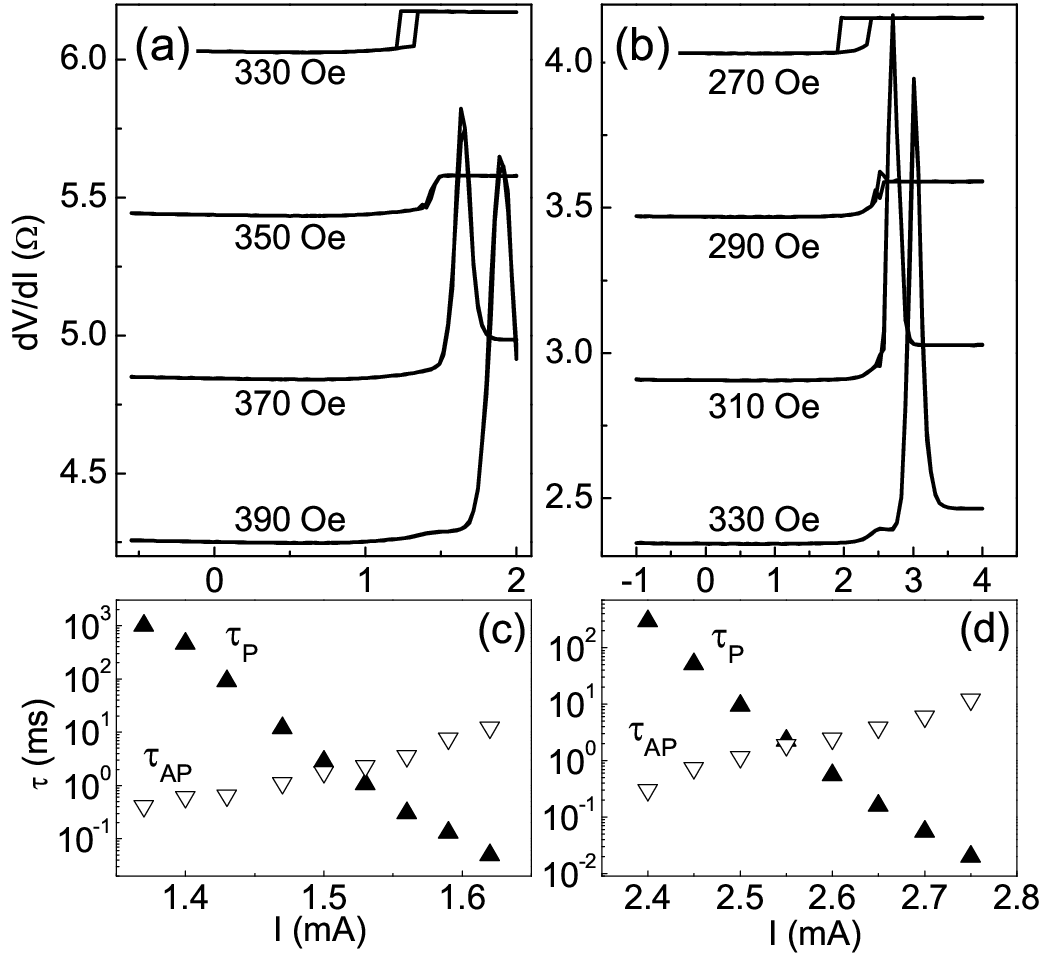}
\caption{\label{fig2} RT data. (a,b) $dV/dI$ for samples $A$, $B$ at the labeled values of $H$. Curves are offset for clarity. (c) Dependence of average dwell times on $I$ in P state (solid symbols) and AP state (open symbols) for sample $A$, at $H=360$~Oe. (d) same as (c), for sample $B$ at $H=285$~Oe.}
\end{figure}

In sample $A$, the current $I^+_S(T)$ for switching from the low-resistance parallel (P) to the high-resistance antiparallel (AP) state increased from $0.90$~mA at RT to $1.23$~mA at $5$~K, with the most significant variation at $T<50$~K (Fig.~\ref{fig1}(a)). The current $I^-_S$ for switching from AP to P state decreased from $-0.38$~mA at RT to $-0.73$~mA at $5$~K. These variations can originate from the reduced thermal activation, changes of spin-dependent transport properties,~\cite{CoPy} or increased damping due to oxidation.~\cite{emley,oxidation} 

At $H=500$~Oe exceeding the coercive field of the nanopillar $H_C=430$~Oe at $5$~K, the $dV/dI$ sharply increased at the onset of current-induced precession $I^+_C=0.92$~mA (Fig.~\ref{fig1}(c)). The $20$~Oe data show a similar increase, indicating large-amplitude precession before reversal at small $H$. The dotted vertical line shows that the onset of precession does not significantly depend on $T$, and coincides with $I^+_S(RT)$.~\cite{mystprl} The same relationship between $I^-_S$ and the precession onset current in the AP state $I^-_C$ is also apparent from the $dV/dI$ data in Fig.~\ref{fig1}(a). Therefore, the increase of $I_S$ at low $T$ can be attributed entirely to thermal effects, with no evidence for increased damping due to nanopillar oxidation. 

\begin{figure}
\includegraphics[width=3.2in]{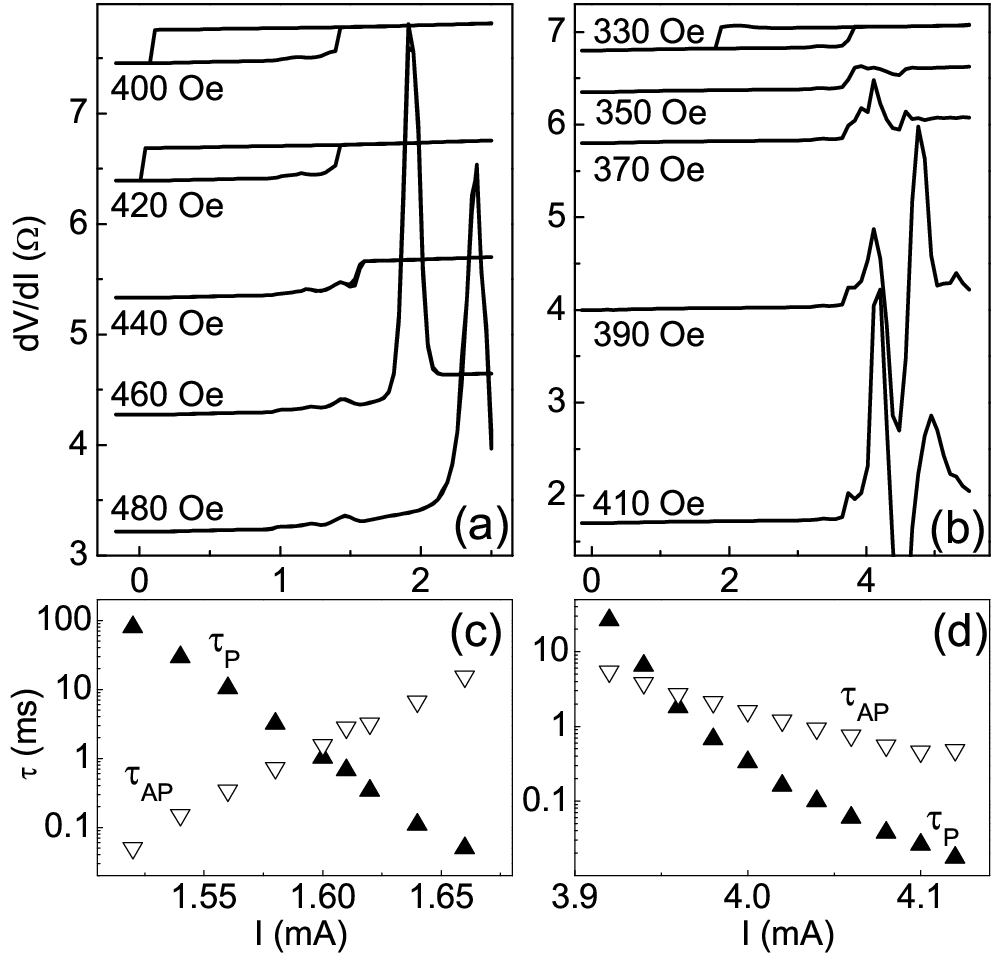}
\caption{\label{fig3} $5$~K data. (a,b) $dV/dI$ for samples $A$, $B$ at the labeled values of $H$. Curves are offset for clarity. (c) Dependence of average dwell times on $I$ in P state (solid symbols) and AP state (open symbols) for sample $A$, at $H=448$~Oe. (d) same as (c), for sample $B$ at $H=370$~Oe.}
\end{figure}

We also compare CIMS in sample $A$ with a nearly identical nanopillar protected by AlO$_x$ coating. From Fig.~2(a) of Ref.~\cite{oxidation}, $I^-_S=-2.0$~mA, $I^-_C=-0.7$~mA, $I^+_S=2.3$~mA, $I^+_C=1.25$~mA at $4.2$~K. The corresponding $5$~K values for sample $A$ are $-0.73$~mA, $-0.38$~mA, $1.23$~mA, and $0.92$~mA. To eliminate the effects of different sample areas, these characteristic currents can be multiplied by the MR $\Delta R=0.23$~$\Omega$ for AlO$_x$-coated sample, and $\Delta R=0.32$~$\Omega$ for sample $A$.~\cite{myiswvsmrapl} The values $(|I|\Delta R)$ for sample $A$ are $0.23$~mV, $0.12$~mV, $0.39$~mV, and $0.29$~mV, with $I=I^-_S$, $I^-_C$, $I^+_S$, and $I^+_C$, respectively. They are similar or smaller than the corresponding values $0.46$~mV, $0.16$~mV, $0.53$~mV, and $0.29$~mV for the AlO$_x$-coated sample, indicating negligible effects of oxidation in sample $A$.

For sample $B$, $I^-_S$ decreased from $-1.05$~mA at RT to $-1.81$~mA at $5$~K, and $I^+_S$ decreased from $-1.05$~mA at RT to $-1.81$~mA at $5$~K (Fig.~\ref{fig1}(c)). The RT values are two times larger than for sample $A$, despite similar values of $\Delta R$ ($0.16$~$\Omega$ for $A$ {\it vs} $0.13$~$\Omega$ for $B$). In contrast to sample $A$, $I^+_C$ increased at low $T$, closely following $I^+_S$ (Fig.~\ref{fig1}(d)). A similar relation between  $I^-_S$ and $I^-_C$ is apparent from the $dV/dI$ curves in Fig.~\ref{fig1}(c), not exhibiting precession before switching, or showing an onset $I^-_C$ very close to $I^-_S$. The $5$~K values $\Delta RI^-_C=0.43$~mV and $\Delta RI^+_C=0.9$~mV for sample $B$ are more than three times larger than for sample $A$, consistent with enhanced low-temperature damping due to oxidation.

To gain further insight into the effects of oxidation, we analyzed reversible CIMS at large $H$ (Figs.~\ref{fig2},~\ref{fig3}). At RT, reversible switching was characterized by large peaks at $H>350$~Oe for sample $A$, and at $H>290$~Oe for sample $B$, which are caused by the thermally-activated random transitions between the P and AP states. The average dwell time $\tau_P$ in the P state decreases, and the average dwell time $\tau_{AP}$ in the AP state increases with increasing $I$ (Figs.~\ref{fig2}(c,d)), resulting in a peak at $\tau_{AP}=\tau_P$.~\cite{mystprl} 

At $5$~K, sample $A$ exhibited a similar reversible switching peak due to the decreasing $\tau_P$ and increasing $\tau_{AP}$ (Figs.~\ref{fig3}(a,c)). In contrast, sample $B$ showed irregular variations of $dV/dI$, but no reversible switching peak (Fig.~\ref{fig3}(b)). Anomalous variations of $dV/dI$ at large $H$ were also seen in naturally oxidized nanopillars.~\cite{oxidation} We clarified these behaviors by measurements of reversal statistics, showing that both  $\tau_P$ and $\tau_{AP}$ decreased with increasing $I$ (Fig.~\ref{fig3}(d)).

Switching statistics can be described by incorporating spin torque into the Fokker-Planck equation, yielding~\cite{lizhang2}
\begin{equation}
\label{tau} \tau(I)=\tau_0exp(E_b[1-I/I_C]/k_BT), 
\end{equation}
where $\tau_0\approx 10^{-9}$~s is the inverse attempt rate, $E_b$ is the activation barrier, and $k_B$ is the Boltzmann constant.  Formula~\ref{tau} is valid only for $I<I_C$, and is not strictly applicable for activation from the P state at $I>I^+_C$ in Figs.~\ref{fig2}(c,d). Nevertheless, it qualitatively describes the observed exponential reduction of $\tau_{P}$ with $I$. For activation from the AP state, Eq.~\ref{tau} with $I_C=I^-_C<0$ predicts an exponential increase of $\tau_{AP}$ with $I$, consistently with RT data for both samples, and $5$~K data for sample $A$. The failure of this formula to describe $\tau_{AP}(I)$ in sample $B$ at $5$~K suggests that reversal occurs over multiple activation barriers. Antiferromagnetic (AF) NiO formed by the oxidation of the nanopillar has a weak magnetic anisotropy, which is generally insufficient to stabilize its magnetic structure in contact with a ferromagnet.~\cite{NiOthermal,stohrspring} Local pinning by the fluctuating AF moments likely produces multiple activation barriers. At large $H$, the activation barrier in the AP state is significantly smaller than in the P state, resulting in larger relative fluctuations of the barrier caused by the same AF pinning, and leading to anomalous behaviors of $\tau_{AP}$.

\begin{figure}
\includegraphics[width=3.1in]{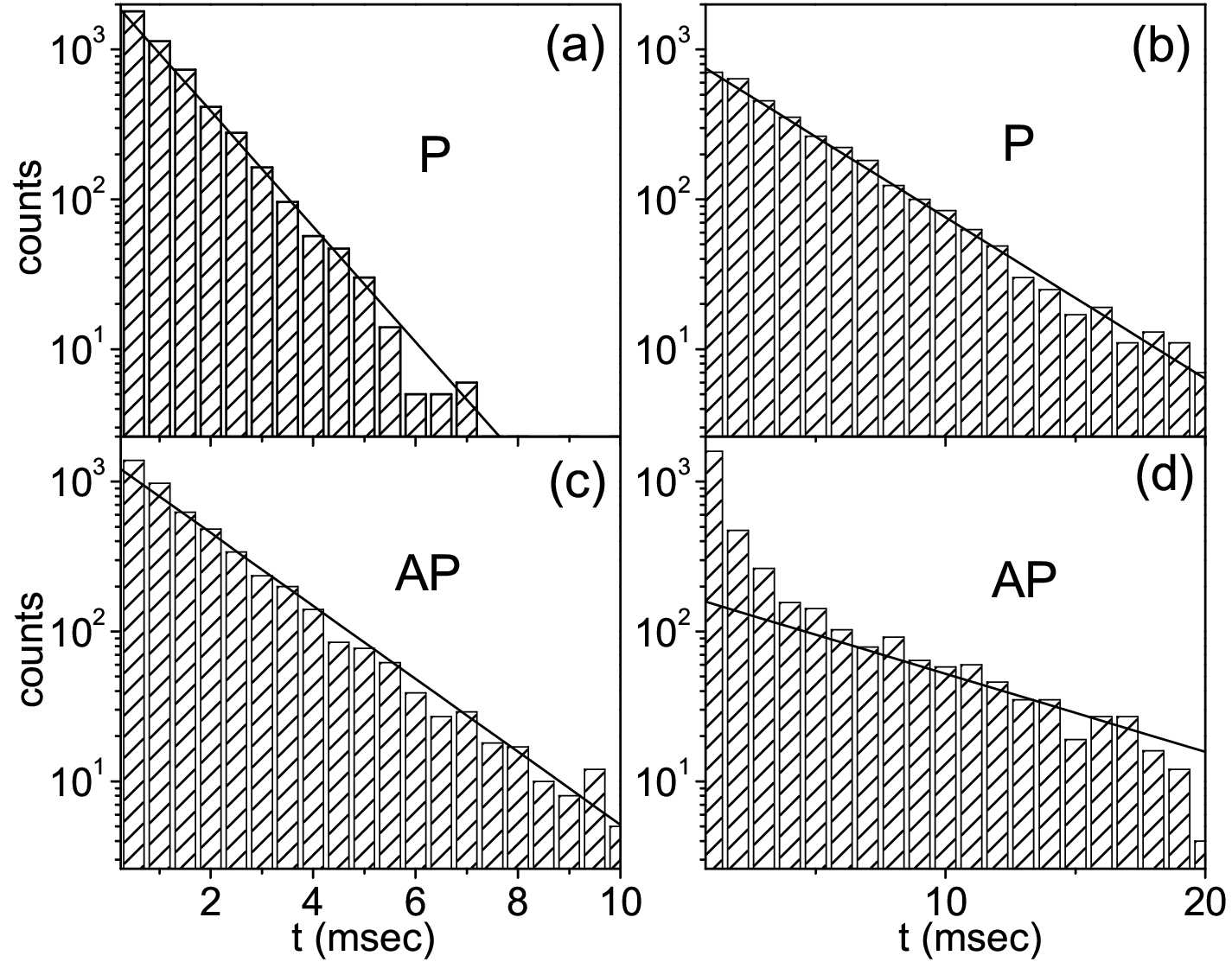}
\caption{\label{fig4} $5$~K data. (a,c) Distribution of dwell times in the P state (a) and AP state (c) for sample $A$ at $H=448$~Oe, $I=1.60$~mA. (b,d) same as (a,c), for sample $B$ at $H=370$~Oe, $I=3.96$~mA. Solid lines: best linear fits on log-linear scale.}
\end{figure}

Both samples at RT, and sample $A$ at $5$~K exhibited an exponential distribution of dwell times in P and AP states, as expected for random thermally activated reversal over a single activation barrier (see Figs.~\ref{fig4}(a,c) for sample $A$ at $5$~K).~\cite{lizhang2} The dwell times of sample $B$ in the P state at $5$~K are also well described by the exponential distribution (Fig.~\ref{fig4}(b)). However, the dwell times of sample $B$ in the AP state do not follow a single exponential distribution, due to the anomalously large number of low-dwell counts (Fig.~\ref{fig4}(d)). This behavior can be attributed to the multiple-barrier activation with at least two significantly different characteristic dwell times, consistent with the fluctuating pinning by the AF oxide discussed above. We note that a similar absence of the reversible switching peak, anomalous dependence of dwell times on $I$, and non-exponential reversal statistics were also observed in nanopillars exchange-biased by a thin AF FeMn, supporting the origin of anomalies in sample $B$ from the AF surface oxide.~\cite{myebstprl}

In summary, we demonstrated that magnetic nanopillars can be efficiently protected from oxidation by sputtered insulating silicon, resulting in excitation onset current nearly independent of temperature. In contrast, oxidized nanopillars exhibit larger switching currents, increased excitation onset currents at low temperatures, and anomalous low-temperature statistics of current-induced reversal. These behaviors provide key signatures for the effects of oxidation, which can be conveniently used in future studies to characterize the quality of magnetic nanodevices. We expect our passivation technique to be transferrable to other nanomagnetic devices such as tunnel junctions.

This work was supported by NSF DMR-0747609 and a Cottrell Scholarship from the Research Corporation. We thank Weng Lee Lim for helpful contributions.


\begin{thebibliography}{99}
\bibitem{slonczewski96} J. Slonczewski, \Journal{J. Magn. Magn. Mater.}{159}{L1}{1996}.
\bibitem{emley} N.C. Emley, I.N. Krivorotov, O. Ozatay, A.G.F. Garcia, J.C. Sankey, D.C. Ralph, and R.A. Buhrman, \Journal{Phys. Rev. Lett.}{96}{247204}{2006}.
\bibitem{oxidation} O. Ozatay, P.G. Gowtham, K.W. Tan, J.C. Read, K.A. Mkhoyan, M.G. Thomas, G.D. Fuchs, P.M. Braganca, E.M. Ryan, K.V. Thadani, J. Silcox , D.C. Ralph , and R.A. Buhrman, \Journal{Nature Mat.}{7}{567}{2008}.
\bibitem{CoPy} S. Urazhdin and S. Button, \Journal{Phys. Rev.}{B 78}{172403}{2008}.
\bibitem{mystprl} S. Urazhdin, N.O. Birge, W.P. Pratt Jr., and J. Bass, \Journal{Phys. Rev. Lett.}{91}{146803}{2003}.
\bibitem{myiswvsmrapl} S. Urazhdin, N.O. Birge, W.P. Pratt Jr., and J. Bass, \Journal{Appl. Phys. Lett.}{84}{1516}{2004}.
\bibitem{lizhang2} Z. Li, and S. Zhang, \Journal{Phys. Rev.}{B 68}{024404}{2003}.
\bibitem{NiOthermal} P.A.A. van der Heijden, T.F.M.M. Maas, W.J.M. de Jonge, J.C.S. Kools, F. Roozeboom, and P.J. can der Zaag, \Journal{Appl. Phys. Lett.}{72}{492}{1998}.
\bibitem{stohrspring} A. Scholl, M. Liberati, E. Arenholz, H. Ohldag, and J. Stohr, \Journal{Phys. Rev. Lett}{92}{247201}{2004}.
\bibitem{myebstprl} S. Urazhdin and N. Anthony \Journal{Phys. Rev. Lett.}{99}{46602}{2007}.
\end{thebibliography}
\end{document}